\documentclass[a4paper,onecolumn,12pt]{quantumarticle}
\pdfoutput=1
\usepackage[english]{babel}
\usepackage[numbers]{natbib}
\usepackage{amsmath,amssymb,hyperref,doi}

\begin{document}

\title{The signaling dimension of physical systems}

\author{Michele Dall'Arno}

\email{dallarno.michele@yukawa.kyoto-u.ac.jp}

\affiliation{Yukawa Institute for Theoretical Physics, Kyoto
  University, Sakyo-ku, Kyoto, 606-8502, Japan}

\affiliation{Faculty  of Education  and Integrated  Arts and
  Sciences, Waseda University, Shinjuku-ku, Tokyo, 169-8050,
  Japan}

\thanks{Report number: YITP-22-129}

\maketitle

\textit{This is a Perspective  on ``Classical simulations of
  communication channels'' by P\'eter E.  Frenkel, published
  in Quantum 6, 751 (2022).}

\vspace{1cm}

The question we start our analysis from is whether a quantum
system  can  outperform  a  classical  system  of  the  same
dimension    for   some    communication   scenarios.     By
``communication scenario'', we mean the usual setup in which
some classical message is encoded  into the system and later
retrieved, although we  do not restrict the  analysis to the
discrimination problem. Such a  question clearly lies at the
heart of quantum communication  theory. Back in 1973, Holevo
gave a  partial answer~\cite{Hol73} by considering  the case
of  asymptotically many  rounds of  communication using  the
quantum system  and proving that the  relevant quantifier of
information  in  this  case,  that  is  the  Shannon  mutual
information, indeed  cannot outperform  that of  a classical
system of the same dimension.

During the subsequent four  decades, researchers followed in
Holevo's  footsteps  by  proving analogous  results  --  the
impossibility for a quantum system to outperform a classical
system of the same dimension in a communication setup -- for
different setups,  and accordingly different  quantifiers of
information.  Among the overwhelming amount of works in this
direction,  it  seems  particularly relevant  to  mention  a
result~\cite{EE07} from 2007 by Elron and Eldar, that can be
considered a precursor of the  breakthrough that was to come
only  a few  years later.   Elron and  Eldar showed  that no
quantum system  can outperform  a classical system  of equal
dimension in  any discrimination  scenario (that is  to say,
technically,   for  linear   games   with  diagonal   payoff
matrix). Still, many communication  setups are not instances
of a  discrimination problem, and  clearly whether or  not a
quantum system can outperform a classical one depends on the
specific case. Or does it?

The anticipated groundbreaking result  is the answer to this
question  in the  most general  case, given  by Frenkel  and
Weiner~\cite{FW15} in  2015.  But  before proceeding,  it is
time to introduce  some notation.  A physical  system $S$ of
linear dimension $\ell \in \mathbb{N}$ can be represented by
a   triple   $(\mathcal{S},  \mathcal{E},   \cdot)$,   where
$\mathcal{S}  \subseteq  \mathbb{R}^\ell$  and  $\mathcal{E}
\subseteq \mathbb{R}^\ell$ are the  set of admissible states
and  effects, respectively,  and  $\cdot$  denotes an  inner
product in  $\mathbb{R}^\ell$. The probability  of measuring
the effect $\pi  \in \mathcal{E}$ given the  state $\rho \in
\mathcal{S}$ is  given by $\rho  \cdot \pi$, and  the effect
that gives  unit probability  for any  state is  called unit
effect.  For   any  $d   \in  \mathbb{N}$,  we   denote  the
$d$-dimensional classical  and quantum systems by  $C_d$ and
$Q_d$,  respectively.   That  is,  for $C_d$  one  has  that
$\mathcal{S}$ is the $(d  - 1)$-dimensional regular simplex,
while for $Q_d$ one has that  $\mathcal{S}$ is the set of $d
\times  d$ (vectorized)  density matrices;  in either  case,
$\cdot$ is  the usual dot  product and $\mathcal{E}$  is the
set  induced  by  the  requirement  that  probabilities  are
non-negative.

Given two  finite alphabets $\mathcal{X}$  and $\mathcal{Y}$
with  cardinality $m$  and  $n$,  respectively, an  encoding
$\boldsymbol{\rho}$   is  a   map   from  $\mathcal{X}$   to
$\mathcal{S}$, while a decoding  $\boldsymbol{\pi}$ is a map
from $\mathcal{Y}$ to $\mathcal{E}$,  such that $\sum_{y \in
  \mathcal{Y}}  \boldsymbol{\pi} \left(  y  \right)$ is  the
unit effect.  We consider  the set $\mathcal{P}^{m \to n}_S$
of  input-output conditional  probability distributions  $p$
that   can  be   generated   by  system   $S$  with   shared
randomness. That  is, $p$  is an element  of $\mathcal{P}^{m
  \to  n}_S$  if and  only  if  there  exists a  family  $\{
\boldsymbol{\rho}_\lambda  \}_\lambda$  of encodings  and  a
family $\{ \boldsymbol{\pi}_\lambda \}_\lambda$ of decodings
such that
\begin{align*}
  p \left(x  | y \right)  = \sum_{\lambda} q  \left( \lambda
  \right)  \boldsymbol{\rho  }_{\lambda}  \left(  x  \right)
  \cdot \boldsymbol{\pi}_{\lambda} \left( y \right),
\end{align*}
for some probability distribution $q$.

We are now  in a position to introduce the  main quantity of
interest   in    our   analysis.     The   \textit{signaling
  dimension}~\cite{DBTBV17} of a physical  system $S$ is the
minimum dimension of a  classical channel that can reproduce
the set of input-output  correlations (or probability range)
attainable by system $S$, that is
\begin{align*}
  \operatorname{sign.dim}  \left( S  \right) :=  \min_{d \in
    \mathbb{N}} d \qquad \textrm{s.t.} \qquad \mathcal{P}^{m
    \to  n}_S  \subseteq  \mathcal{P}^{m  \to  n}_{C_d},  \;
  \forall m, n \in \mathbb{N}.
\end{align*}
Clearly, whether  or not there exists  a communication setup
in  which  a  quantum  channel can  outperform  a  classical
channel of  equal dimension can be  conveniently reframed in
terms  of  the  signaling  dimension;  in  this  sense,  the
signaling dimension  summarizes the structure of  the entire
set of  input-output correlations that is  consistent with a
given system in a single  scalar quantity.  This is in stark
contrast  with  previous   approaches  addressing  the  same
problem that  were based on  specific choices of  a witness,
that  is, that  investigated the  correlation space  along a
single direction only.

Equipped with the definition  of the signaling dimension, we
can now announce the aforementioned breakthrough~\cite{FW15}
achieved by  Frenkel and Weiner  in 2015, that  consisted in
proving that
\begin{align*}
  \operatorname{sign.dim} \left( Q_d \right) = d, \; \forall
  d \in \mathbb{N},
\end{align*}
or, in words, no quantum  channel can outperform a classical
channel  of equal  dimension  in \textit{any}  communication
game. Frenkel  and Weiner  obtained this powerful  result by
adopting graph theoretic  techniques and mixed discriminants
that, to  the best of  our knowledge, were a  novelty within
quantum information  theory at that time.   As a consequence
of their  remarkable finding, in  the same work  the authors
also derived  a combinatorial  Holevo-like bound  that turns
out  to  be tighter  that  the  original  in at  least  some
scenarios.

Once it became clear that  quantum systems are equivalent to
classical systems  of equal  dimension in  any communication
setup,  the question  immediately arose  how unique  quantum
theory  is  in  this  respect within  the  set  of  physical
theories  or,  more  accurately,  generalized  probabilistic
theories. Since  for arbitrary  systems -- that  is, systems
other than  classical or quantum  -- there is no  concept of
``dimension'',  to  answer  this  question  a  criterion  to
meaningfully compare systems belonging to different theories
had to be  derived. This was achieved by  observing that, by
definition, the  signaling dimension  of the  composition of
multiple classical systems  is just given by  the product of
the   signaling  dimensions   of   the  component   systems.
Informally,  this is  to  say that  using classical  systems
sequentially allows for the  same amount of communication as
using them in parallel. As a formula this statement reads
\begin{align*}
  \operatorname{sign.dim} \left( S_0 \otimes S_1 \right) \le
  \operatorname{sign.dim}      \left(       S_0      \right)
  \operatorname{sign.dim} \left( S_1 \right),
\end{align*}
where  $\otimes$ denotes  the  composition  of systems  (not
necessarily  given by  the tensor  product). This  seemingly
obvious    fact,     formally    given    the     name    of
\textit{no-hypersignaling   principle}~\cite{DBTBV17},  also
holds true for quantum theory,  but only as a consequence of
the aforementioned result~\cite{FW15} by Frenkel and Weiner,
and provides the promised  criterion to meaningfully compare
systems belonging to different theories.

It  must be  remarked that  the no-hypersignaling  principle
constrains  the  time-like  correlations that  any  physical
theory  can exhibit.   This is  in contrast  with previously
known  principles,  including   the  no-signaling  principle
itself, which instead bounds  the space-like correlations of
any given  physical theory.  Therefore, it  is remarkable to
observe  that  there   exist  theories~\cite{DBTBV17}  that,
although  equivalent to  quantum  theory in  terms of  their
space-like  correlations,  exhibit  super-quantum  time-like
correlations   only   detectable   as  violations   of   the
no-hypersignaling  principle.   Moreover,  it  was  recently
shown~\cite{DEP20a, DEP20b} by D'Ariano, Erba, and Perinotti
that  there exist  theories that  are locally  classical but
nevertheless violate the no-hypersignaling principle.

From the  resource theoretical point of  view, the signaling
dimension of a  channel -- a system can be  regarded in this
sense as the identity channel -- clearly quantifies the cost
of  classically simulating  such a  channel. The  problem of
quantifying the  signaling dimension of a  given channel has
recently  drawn  considerable  attention. While  a  detailed
analysis  of the  literature  is outside  the  scope of  the
present perspective, here  we just mention a  few works: for
the case of  quantum channels, the works  by Hoffmann, Spee,
G\"uhne, and Budroni~\cite{HSGB18}, Dall'Arno, Brandsen, and
Buscemi~\cite{DBB20},  Doolittle and  Chitambar~\cite{DC21},
and Chitambar,  George, Doolittle,  and Junge~\cite{CGDJ21};
for  entanglement-assisted classical  channels, the  work by
Frenkel   and  Weiner~\cite{FW22};   and  for   channels  in
generalized  probabilistic  theories,  the  recent  work  by
Heinosaari, Kerppo, and Leevi Lepp\"aj\"arvi~\cite{HKL20}.

Ultimately,  the  signaling dimension  is  the  result of  a
comparison between a  given generalized probabilistic theory
-- possibly quantum theory --  and classical theory, through
their  probability  ranges.   A  context  in  which  such  a
comparison     is    inherently     meaningful    is     the
device-independent    approach   to    quantum   information
processing, in which classical theory  plays the role of the
standard  reference  against  which   any  other  theory  is
compared.  This is because  classical events -- the pressing
of a  button, the detection of  a light bulb lighting  up --
are assumed  as the only  events directly accessible  by the
observer.

Therefore, it  comes as no surprise  that device-independent
tests of  quantum channels based on  the signaling dimension
have been proposed~\cite{DBB17, DBBV17, Dal19}. Therein, the
probability range of a given quantum device is compared with
observations in order  to rule out, in  a device independent
way, that such  a device has produced  the observations.  In
this  context,   Frenkel  and   Weiner's  result~\cite{FW15}
established that, unless one abandons the input-output setup
-- as  done for  instance by  Gallego, Brunner,  Hadley, and
Ac\`in~\cite{GBHA10} --, the  identity channels in classical
and quantum theories  remain indistinguishable.  These ideas
were further  pushed forward~\cite{BD19, DBBT20,  DHBS19} by
the same  authors in  collaboration with Bisio,  Tosini, Ho,
and Scarani,  by developing  data driven techniques  for the
inference of quantum devices.

Perhaps,   one  of   the  most   promising  research   lines
originating  from the  study of  the signaling  dimension is
based  on an  attempt to  unify the  aforementioned resource
theoretical  and  device-independent approaches  to  quantum
information  theory.   To  this  end,  recently,  Dall'Arno,
Buscemi, and Scarani laid  the foundations~\cite{DBS20} of a
device-independent approach  to resource theory, that  is, a
scenario  in  which  operationally  meaningful  majorization
relations between resources are  established on the basis of
observed input-output correlations only.

The  new  paper~\cite{Fre22}  by  Frenkel,  for  which  this
perspective is  written, is a  timely addition to  this rich
research  landscape.    Therein,  the  author   extends  the
groundbreaking  result~\cite{FW15}  he   obtained  almost  a
decade  ago  in  collaboration  with  Weiner,  thus  further
pushing   the  state   of  the   art  in   the  problem   of
characterizing the  signaling dimension. The  paper provides
bounds  on  the  signaling  dimension in  terms  of  another
quantifier   of  information,   that  is,   the  information
storability~\cite{MK18}.    The    information   storability
quantifies the  tradeoff between the success  probability in
the state discrimination  and the size $m$  of the alphabet;
as a formula:
\begin{align*}
  \operatorname{inf.stor}     \left(     S    \right)     :=
  \max_{m, \boldsymbol{\rho},  \boldsymbol{\pi}}   \sum_{k  =
    1}^{m}     \boldsymbol{\rho}     \left(    k     \right)
  \boldsymbol{\pi} \left( k \right),
\end{align*}
where the optimization is over any size $m$ and any encoding
and decoding.   While previous  literature on  the signaling
dimension in  general probabilistic theories  mostly focused
on  systems whose  state  space is  a  regular polygon,  the
bounds  derived  by  Frenkel  can be  used  to  compute  the
signaling  dimension  of   regular  polyhedrons.   Moreover,
Frenkel  also shows  that, when  the state  space is  a ball
according to the $k/(k-1)$-norm,  the signaling dimension is
upper bounded  by $k$, and provides  a full characterization
of  the  signaling dimension  of  several  classes of  noisy
quantum channels.

But perhaps  the most significant contribution  of Frenkel's
work~\cite{Fre22}   lies  in   having  pushed   forward  the
mathematical  techniques, based  on graph  theory and  mixed
discriminants,       on       which       his       previous
breakthrough~\cite{FW15}, in collaboration  with Weiner, was
based.   Recently,  these techniques  were  the  topic of  a
course, held by  Frenkel and Weiner, at the  Kyoto School on
Advanced  Topics  in  Quantum Information  and  Foundations,
whose lectures are  freely available online~\cite{FW21}.  It
is hard not to wonder about the impact these techniques will
have in quantum information theory in the years to come.

\bibliography{sigdim}{}
\bibliographystyle{unsrtnat}
\end{document}